\documentclass[12pt,preprint]{aastex}

\shorttitle{Slow magnetoacoustic oscillations in the microwaves}
\shortauthors{Kim, Nakariakov, \& Shibasaki}

\begin{document}

\title{Slow Magnetoacoustic Oscillations in the Microwave Emission of Solar Flares}

\author{S. Kim}
\affil{Nobeyama Solar Radio Observatory / NAOJ, Nagano 384-1305,
Japan} \email{sjkim@nro.nao.ac.jp}

\and

\author{V.~M. Nakariakov}
\affil{Physics Department, University of Warwick, Coventry, CV4 7AL, UK}
\affil{Central Astronomical Observatory of the Russian Academy of Sciences at Pulkovo, 196140 St Petersburg, Russia}

\and

\author{K. Shibasaki}
\affil{Nobeyama Solar Radio Observatory / NAOJ, Nagano 384-1305, Japan}

\begin{abstract}
Analysis of the microwave data, obtained in the 17 GHz channel of
the Nobeyama Radioheliograph during the M1.6 flare on 4th Nov
2010, revealed the presence of 11.8-min oscillations of the
emitting plasma density. The oscillations decayed with the
characteristic time of about 25-min. These oscillations are also
well-seen in the variation of EUV emission intensity measured in
the 335\, \AA\ channel of SDO/AIA. The observed properties of the
oscillations are consistent with the properties of so-called SUMER
oscillations, observed in the EUV and soft X-ray bands usually as
a periodic Doppler shift. The accepted interpretation of SUMER
oscillations is a standing slow magnetoacoustic wave. Our analysis
presents the first direct observation of the slow magnetoacoustic
oscillations in the microwave emission of a solar flare.
\end{abstract}


\section{Introduction}

Quasi-periodic pulsations (QPP) in the emission generated in solar flares, with the periods ranging
from a fraction of second to several minutes, have been intensively studied
for several decades \citep[see, e.g.][for a recent review]{2009SSRv..149..119N}. One of the possibilities
opened up by revealing the nature of QPP in flares is the diagnostics of physical conditions in
flaring sites and mechanisms operating in them. Moreover, this diagnostics can be extended to stellar flares,
which are also observed to have QPP in their radio, optical and soft X-ray light curves
\citep[e.g.][]{2001A&A...374.1072S,2003A&A...403.1101M,2005A&A...436.1041M}.
The origin of the QPP is still not clear, while it is accepted that they can be produced by several
mechanisms. Perhaps, one of the most understood possibility is the generation of QPP by
magnetohydrodynamic (MHD) oscillations of coronal structures. Coronal MHD oscillations are well seen in
various bands directly with the modern high time and spatial resolution instruments, which
provide researchers with the ground for their identification in the flaring emission.

Standing longitudinal oscillations of coronal loops were first detected as
the periodic Doppler shift of the emission lines Fe\,{\sc xix} and
Fe\,{\sc xxi}, with the formation temperature greater than 6 MK,
with the SOHO/SUMER instrument (\citealt{2003A&A...402L..17W};
\citealt{2003A&A...406.1105W}). The mean observed oscillation period
is 17.6$\pm$5.4 min. The
oscillations are strongly damped, with the damping time about one
period of the oscillation. In some cases, the intensity
oscillations are seen. There is a quarter-period phase lag between
the intensity and the Doppler shift oscillations. The oscillations
are usually observed in association with the soft X-ray
brightenings, sometimes up to the M-class flares
\citep{2007ApJ...656..598W}. Similar Doppler-shift oscillations
during solar flares in emission lines of S\,{\sc xv} and Ca\,{\sc
xix}, with the formation temperature 12--14 MK, with Yohkoh/BCS
were reported by \cite{2006ApJ...639..484M, 2005ApJ...620L..67M}.
In cooler coronal emission lines, similar oscillations were
detected with Hinode/EIS \citep{2008ApJ...681L..41M}.

The compressible nature of the longitudinal oscillations and their
long periods led to their interpretation in terms of standing slow
magnetoacoustic oscillations damped because of high thermal
conduction \citep{2002ApJ...580L..85O}. A series of numerical
studies \citep[e.g.][]{2004A&A...414L..25N, 2004ApJ...605..493M,
2004A&A...422..351T, 2005A&A...436..701S, 2005A&A...438..713T,
2007A&A...467..311O, 2007ApJ...668L..83S,2011A&A...536A..68G}, accounting for various
additional physical effects including viscosity, 2D and 3D
geometry, stratification, nonlinear steepening and mode coupling,
demonstrated the robustness of this interpretation. In particular,
the simulations showed that, depending upon whether the
oscillations are triggered at one or both footpoints of a coronal
loop, the fundamental mode and its second spatial harmonics can be
effectively excited. The modes have different structure of the
oscillations at the loop apex: in the fundamental mode there is a
node of the density perturbation and the maximum of the
field-aligned velocity perturbation at the apex, while it is the
other way around in the second harmonics. The phase speed of the longitudinal waves is
the tube speed $C_T=C_s C_A/\sqrt{C_s^2+C_A^2}$, where $C_s$ and $C_A$ are the sound and Alfv\'en speeds,
respectively. The tube speed is
subsonic and sub-Alfv\'enic. In a low-beta plasma the tube speed is just slightly lower
than the sound speed, while in the case $C_s=C_A$ the tube speed decreases to about $0.7C_s$.
With the decrease in the Alfv\'en speed $C_A<C_s$ the tube speed remains lower than the Alfv\'en speed.

Slow magnetoacoustic oscillations can be also observed in the
non-thermal emission of solar and stellar flares. The possible
mechanisms are the periodic triggering of magnetic reconnection by
the modulation of the physical conditions in the vicinity of the
reconnection site \citep{2006SoPh..238..313C}, the modification of
the spectral maximum of the gyrosynchrotron emission in the regime
of the Razin suppression because of the periodic modulation of the
electron plasma frequency \citep{2006A&A...446.1151N}, and the
direct modulation of the emission intensity because of the change
of the concentration of the emitting plasma. However, so far these
theoretical possibilities have been without observational
confirmation. In this Letter, we demonstrate, for the first time,
the presence of longitudinal oscillations in the free-free
microwave emission in a solar flare.

\section{Observations}
We have examined the flare loops produced by a M1.6 flare which
occurs at south-east limb on 2010 Nov 4th. For this study, we used
17 GHz and 34 GHz data observed with Nobeyama Radioheliograph
(NoRH; \citealt{1994Proc. of the IEEE...82..701, tak97}) and 335
\AA\ EUV data observed with Atmospheric Imaging Assembly (AIA;
\citealt{lem11}) onboard Solar Dynamics Observatory. Microwave
data has a spatial resolution of 10 \arcsec\ and a time cadence of
10 seconds. AIA data has a spatial resolution of 1.2 \arcsec\ and
a 12 seconds time cadence. AIA 335 \AA\ channel contains a
contribution from Fe\,{\sc XVI} with the peak forming temperature
of 2.5 MK for the flare \citep{lem11}.

Figure 1 shows time profiles of the flare in soft X-rays taken
with GOES satellite (top) and radio fluxes at 17 and 34 GHz
(bottom). The flare starts at 23:55, peaks at 23:57 UT, and then
gradually decays. The fluxes at 17 and 34 GHz vary with same
pattern during flare process but the difference of scale between
them clearly changes as follows: for the flare peak, 17 GHz flux
is larger than 34 GHz and then decrease less than 34 GHz, and
finally both fluxes become comparable with each others. It implies
that only in the flare peak the radio emission is generated by
gyrosynchroton motion of accelerated electrons, while in the long
decay phase, thermal free-free dominate for the radio emission.
Nobeyama Radiopolarimeter (NoRP; \citealt{1985PASJ...37..163}),
that observes the Sun in the 1, 2, 3.75, 9.4, 17, and 35 GHz
channels, supports it with the same aspect at the high frequencies
where the plasma becomes optically thin.

In Figure 2, we present spatial features of radio emission sources
associated with the flare loops. Figure 2a exhibits the AIA 335
\AA\ image taken at 00:30 UT and the brightness temperature
($T_B$) contours of NoRH 17 and 34 GHz. Figure 2b shows NoRH 17
GHz $T_B$ map with a contour of EUV flare loops observed by AIA
335 \AA\ channel. The flare loops are coincided with the microwave
source. The top of the flare loops emanate strong EUV emission
continuously through the whole flare procedure. In Figure 3, we
plotted the variation of the maximum counts recorded in the AIA
335 \AA\ channel for the flare region. Since the top of the flare
loops is the dominant source during the flare, this plot reflects
the intensity variation of the flare loop-top. Interestingly, it
shows an obvious decaying oscillation pattern after the flare peak
time. Unfortunately, we couldn't confirm it in AIA 131 and 94 \AA\
channel, where the peak of the temperature response is higher than
7 MK, due to the saturation effect on the top of the flare loops.

The NoRH observation at two frequencies allows us to derive a
spectral index $\alpha$ for a spatially resolved region such as
the loop top or loop footpoint. The index $\alpha$ is given by the
ratio of the fluxes obtained at 17 and 34 GHz, $\alpha
=\log({F}_{17GHz}/{F}_{34GHz})/\log({17}/{34})$. For the top of
the arcade (the white box in Figure 2b), the derived index
$\alpha$ is close to 0 within range of -0.2 $\sim$ 0.7 during the
decay phase. It imply that the radio emission in this region comes
from an optically thin plasma
\citep{1985Ann.Rev.Astron.Astrophy...23..169}.

We estimated the plasma temperature of the flare loops using the
ratio of the GOES two channels (1--8 \AA\ and 0.5--4 \AA\ ).
Although GOES satellite observes the full Sun, it is reasonable to
use it in our analysis because only this flaring loop is the
predominant source for the increasing X-ray flux during the flare
time. Using GOES widget, we derived the temperature in the flare
loops \citep{whi05}. The temperature response function of each
filter is derived using the coronal abundances in CHIANTI 6.0.1
\citep{der97,der09}. The estimated peak temperature is 18 MK.
Then, the temperature gradually decreases to about 7 MK (top panel
of Figure 3). Since we are focusing on the top of the flare loops
that appears in microwave imaging data, we do not use GOES
emission measure estimated over the whole flare.

\section{Electron Density}
\label{sec_dens} To derive the electron number density of the
flare loop-top region, we used the brightness temperature obtained
at 17 GHz of NoRH and the estimated plasma temperature by the GOES
two channels. Based on the observations, we assume that in the top
of the flare loops the free-free emission from the optically thin
thermal plasma (the optical depth ${\tau}_{\nu} \ll 1$) is
dominant. According to the radiative transfer at the radio
frequency \citep[e.g.][]{1985Ann.Rev.Astron.Astrophy...23..169},
the relation between the brightness temperature $T_B$ and the
plasma temperature $T$ is given as
\begin{equation}
T_B = T {\tau}_{\nu},
\end{equation}
where $\tau_\nu$ is
\begin{equation} {\tau}_{\nu} = \frac{9.786
\times {10}^{-3}}{\nu^2} \ ln \Lambda \ \frac{EM}{T^{3/2}}.
\end{equation}
$EM$ is the emission measure ($EM = \int_{z} N^2 dz$) and $\log
\Lambda$ is the Coulomb logarithm ($\log \Lambda = \log [4.7
\times 10^{10} (T/\nu)]$). The path on the line-of-sight ($z$) is
assumed as the half of the distance between footpoints of the
flare loops seen in the contour on Figure 2b ($z$ = 25 Mm).
Substituting these observational results to equation (2), we
derived $EM$ and then estimated the number density of electrons at
the top of the flare loops. The middle panel of Figure 4 shows the
resulted electron density variation with time during the whole
flare process. Since the peak emission of the flare is not caused
by the thermal free-free but by the strong gyrosynchrotron
emission, the density estimation related with the flare peak time
cannot be analyzed as a physical parameter in our assumptions.
Therefore we have examined the density estimated from 00:00 UT
(the vertical dashed line in Figure 4) to 00:50 UT, from the time
when the emission only come from the free-free transition to just
before the next flare start. From 00:00 UT, the density
fluctuations come into sight and its amplitude gradually diminish
with time.

\section{Decaying oscillations}

The time dependence of the electron density derived by the radio
emission (middle panel of Figure 4), as well as the EUV emission
(Figure 3), shows a decaying long-period oscillatory pattern. To
emphasise the oscillatory pattern we de-trend the observational
curve by subtracting the same curve averaged by 500 s.
Best-fitting the de-trended observational curve from 00:00 UT by a
decaying harmonic function gives us the parameters of this
oscillation: the period of 710 s and the decay time of 1500 s. The
amplitude of the oscillation is about 6\% of the background in the
beginning of the oscillations. The result of this approximation,
shown in the bottom panel of Figure 4, demonstrates a rather good
agreement  between the observational and the best-fitted curve.
Also, we see that the amplitude of short-period fluctuations from
this decaying harmonic curve decreases with time. We would like to
stress that the detection of the oscillation is not sensitive to
the specific choice of the time scale in the boxcar de-trending,
as the oscillatory pattern is clearly visible in the original
signal.

\section{Discussion}

We have demonstrated that there is an electron density QPP
oscillation with 11.8-min period and 25-min decay time, and the
amplitude of several percent of the background in the flare loops.
This electron density have been deduced by radio emission observed
by 17 GHz Nobeyama Radioheliograph and it is first evidence of
radio observation for the long-period oscillation of the thermal
emission produced by a flare. The QPP is also well-pronounced in
the EUV 335\,\AA\ signal obtained with SDO/AIA. The observed
parameters of the oscillation, the period and decay time, as well
as its compressive nature, are similar to the compressive
oscillations of coronal loops, known as SUMER oscillations.

Following the interpretation of SUMER oscillations as a standing
slow magnetoacoustic wave \citep{2002ApJ...580L..85O} we estimate
parameters of the observed oscillation. Consider two lowest
spatial harmonics. In the slow wave the density is perturbed
because of the spatial redistribution of the matter mainly along
the magnetic field. In a loop it corresponds to the field-aligned
movement of the plasma from one footpoint to the other (in the
fundamental mode) or from both the footpoint to the apex (in the
second standing harmonics). Taking  the loop height of about
50\arcsec\ and assuming that the loop is of semi-circular shape,
we get its length of about 115 Mm. The fundamental mode has the
wavelengths of  $\lambda_1\approx 2\times115$~Mm. The wavelength
of the second spatial harmonics is $\lambda_2\approx 115$~Mm.
Estimating the phase speeds as the ratio of the wavelengths to the
value of the period, 710 s, we obtain about 320 km/s for the
fundamental mode and 160 km/s for the second spatial harmonics.
The plasma temperature associated with the 335\,\AA\ channel is
about 2.5 MK, which gives us the sound speed of about $C_s\approx
240$~km/s. This value is consistent with the phase speed required
for the interpretation of the oscillations in terms of the
fundamental acoustic mode \citep[see,
e.g.][]{2007ApJ...656..598W}.

On the other hand, temperature diagnostics performed with the use
of GOES data shows that the plasma has higher temperature, about
7~MK (see Sec.~\ref{sec_dens}). For this temperature, the sound
speed $C_s$ is greater than 380 km/s. In this case, the decrease
in the phase speed can be attributed to the relatively low value
of the Alfv\'en speed $C_A$, of the value of the sound speed. This
reduces the phase speed of the longitudinal waves $C_T$ to the
required value if $C_A \approx C_s$. This result is consistent
with the estimations of the plasma beta
\citep{2001ApJ...557..326S,2008PFR.....2S1012S}, which show that
in flaring loops this parameter could reach or even exceed unity,
giving us the tube speed of the required value. The fact that the
oscillations are well seen in the 335\AA\ channel can be
attributed to the effect of the crosstalk from the 131 \AA\
channel \citep{lem11}. The oscillation is not seen in hotter
channels of AIA because of the saturation.

Also, if the oscillation of
the emission detected in the 335\AA\ channel is produced by the cooler
plasma, this oscillation can be induced by the slow magnetoacoustic
oscillations in the hotter loops. One possible scenario is based upon the
possibility that hot and cool loops form loop bundles, with the steep
temperature gradient in the transverse direction \citep[e.g.][]{2003A&A...404L...1K}.
Slow magnetoacoustic oscillations cause not only longitudinal, but also
transverse motions of the plasma \citep{2011A&A...536A..68G}. These transverse
motions grow with the increase in beta, and are significant when beta is about unity.
Hence, standing slow magnetoacoustic oscillations channeled by hotter loops
would naturally affect the cooler plasma in the cooler loops situated nearby, and
hence may be visible in the cooler temperature emission.

%However in the considered case,
%the density perturbations are observed in the vicinity of the loop top, where the perturbations
%produced by the fundamental mode are zero . Thus, a more
%preferable interpretation would be the second spatial harmonics \citep{2004A&A...414L..25N}.

Thus, we conclude that our results give the first direct observational evidence of
slow magnetoacoustic waves in the solar corona in the thermal radio emission, similar to
the phenomenon known as SUMER oscillations.

The observed gradual decrease in the electron density fluctuations from the
decaying harmonic pattern can be attributed to the decay of the compressible
turbulence in the flaring site, excited by the flare, and should be subject to further investigation.

\acknowledgments
The work was supported by t...

\clearpage

\begin{figure}
\centering
\includegraphics[width=14cm]{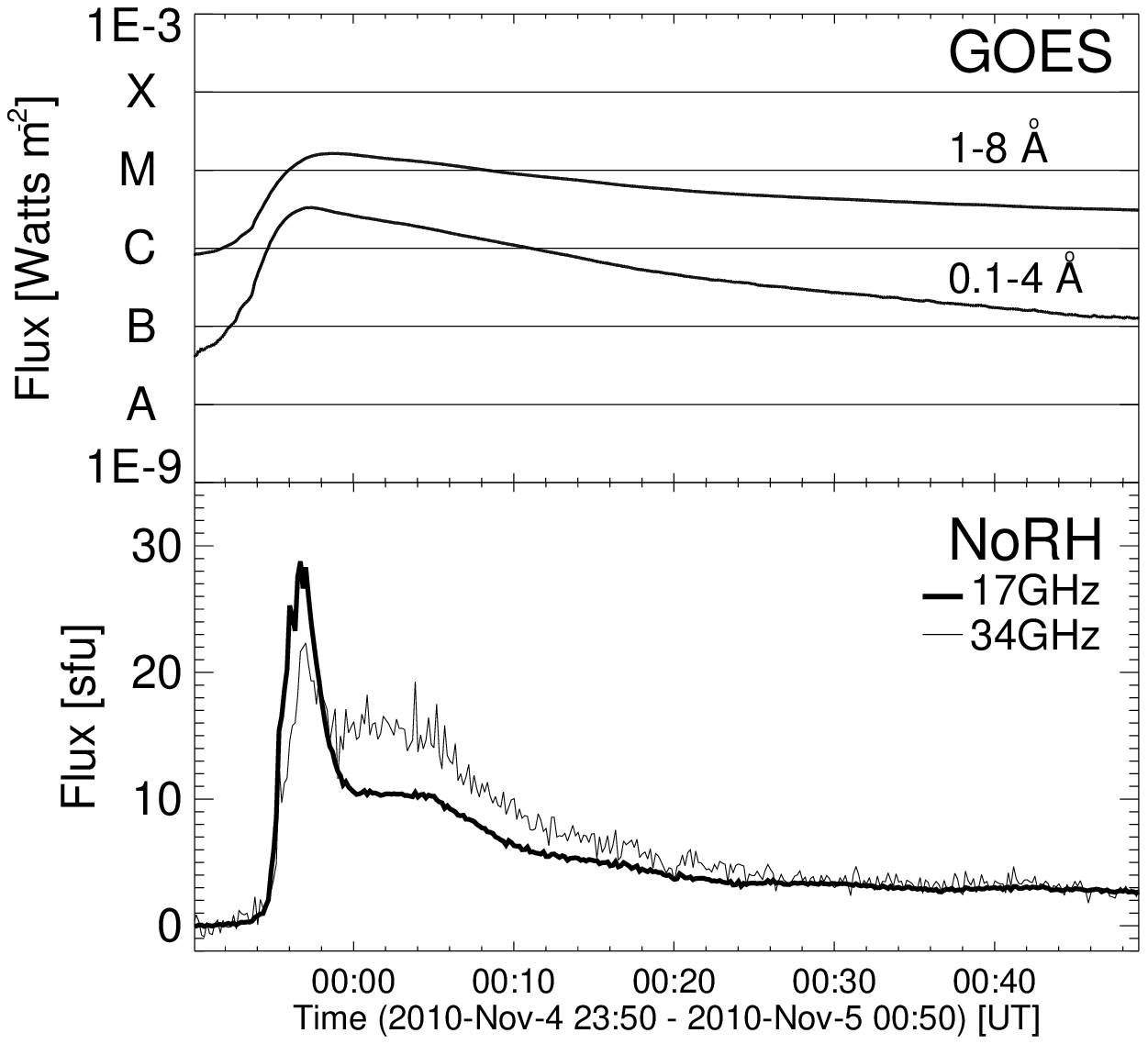}
\caption{Time profile of GOES X-ray (top) and NoRH 17 and 34 GHz
(bottom) flux during a M1.6 flare occurred in 2010 Nov 4th. NoRH
flux at each frequency is estimated over the field of view of
Figure 2.}\label{Fig1}
\end{figure}

\begin{figure}
\centering
\includegraphics[width=14cm]{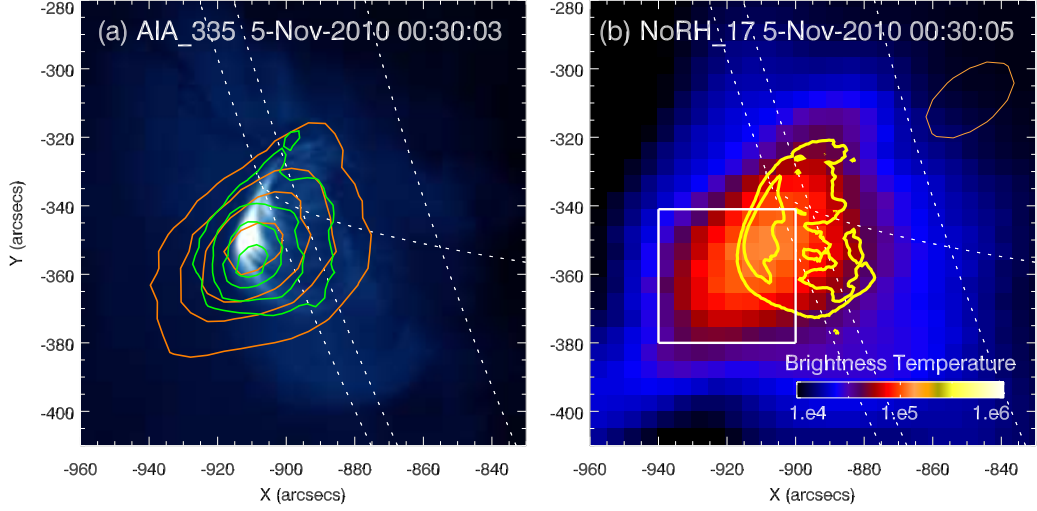}
\caption{Images for the flare loops at around 00:19 UT. (a) AIA
335 \AA\ channel image with contours of $T_B$ at NoRH 17 GHz
(orange) and 34 GHz (green). The levels of contours are 35, 55,
75, and 95 \%\ of peak $T_B$ for each frequency. (b) NoRH 17 GHz
$T_B$ map with contour of flare loop arcade observed by AIA 335
\AA\ channel. A beam size at 17 GHz appears in a yellow circle and
a white box indicates the region where the electron density is
estimated. (A color version of this figure is available in the
online journal.)}\label{Fig2}
\end{figure}

\begin{figure}
\centering
\includegraphics[width=14cm]{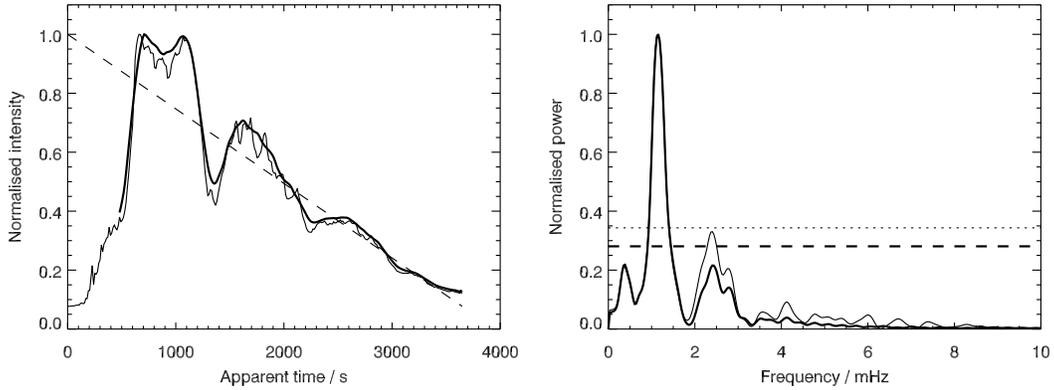}
\caption{Time profile of maximum counts obtained by AIA 335 \AA\
channel in the flare loops}\label{Fig3}
\end{figure}

\begin{figure}
\centering
\includegraphics[width=14cm]{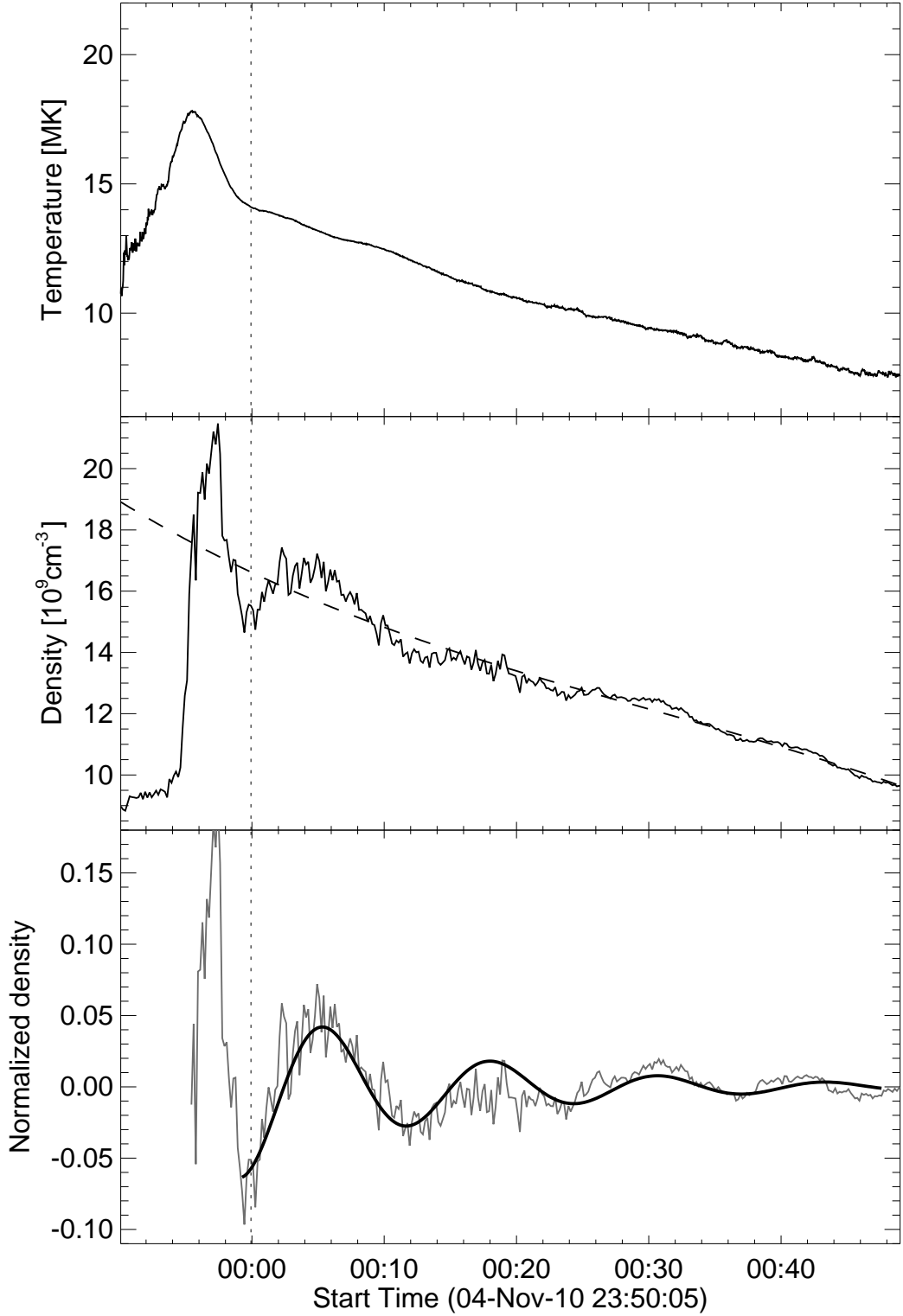}
\caption{Time profile of temperature estimated by GOES two
channels (top), electron number density derived by this
temperature and $T_B$ at 17 GHz (mid), and a best fit on the
normalized electron number density (bottom). Free-free emission
become predominant than gyrosynchrotron emission from 00:00 UT
(vertical dashed-line).}\label{Fig4}
\end{figure}
\end{document}